\documentclass[10pt,twoside]{article}

\usepackage{asp2010}
\usepackage{booktabs}
\usepackage{tabularx}

\aspcpryear{2013}

\aspvolume{467} 
\aspcpryear{2013} 
\aspvoltitle{9\textsuperscript{th} LISA Symposium, Paris} 
\aspvolauthor{G. Auger, P. Bin´etruy and E. Plagnol, eds.} 

\resetcounters
\begin{document}

\resetcounters

\title{Breadboard model of the LISA phasemeter}
 \author{
 	 O.~Gerberding$^1$,
 	 S.~Barke$^1$,
 	 I.~Bykov$^1$,
   K.~Danzmann$^1$, 
   A.~Enggaard$^2$, 	 
   J.J.~Esteban$^1$,
   A.~Gianolio$^3$,   
   T.V.~Hansen$^2$,    	 
   G.~Heinzel$^1$,
   A.~Hornstrup$^4$,
   O.~Jennrich$^3$,
   J.~Kullmann$^1$,
   S.M.~Pedersen$^4$,      	 
   T.~Rasmussen$^2$,
   J.~Reiche$^1$, 
   Z.~Sodnik$^3$
   and~M.~Suess$^3$
 \affil{$ö1$ Max-Planck Institute for Gravitational Physics (Albert Einstein Institute) and Institute for Gravitational Physics, Leibniz University Hannover, Callinstrasse 38, 30167 Hannover, Germany}
 \affil{$ö2$ Axcon ApS, Diplomvej 381, DK-2800 Kgs. Lyngby, Denmark}
 \affil{$ö3$ European Space Research and Technology Centre, European Space Agency, Keplerlaan 1, 2200 AG Noordwijk, The Netherlands}
 \affil{$ö4$ DTU Space, National Space Institute, The Technical University of Denmark, Elektrovej 327, DK-2800 Kgs. Lyngby, Denmark}
 }
 
\begin{abstract} 
An elegant breadboard model of the LISA phasemeter is currently under development by a Danish-German consortium. The breadboard is build in the frame of an ESA technology development activity to demonstrate the feasibility and readiness of the LISA metrology baseline architecture. This article gives an overview about the breadboard design and its components, including the distribution of key functionalities.
\end{abstract}

\section{Introduction} 

One of the main building blocks of the LISA metrology system is the frequency distribution and phase measurement system (phasemeter) (\cite{Jennrich2009}). We are currently developing a breadboard model (BB) of the LISA phasemeter in the frame of an ESA technology development activity. The BB will be used to test and validate the  functionality and performance as required for the LISA metrology system. We will demonstrate the technological readiness and evaluate a future space qualification. The activity is performed by a consortium composed of the National Space Insitute of the Danish Technical University (DTU Space), the Danish industry partner Axcon ApS in Copenhagen, and the Max-Planck Institute for Gravitational Physics, Albert-Einstein Institute (AEI) in Hannover.

This activity is part of a world wide effort to develop and demonstrate the metrology concepts for the LISA interferometry (\cite{Spero2011,Mitryk2010, Heinzel2011}). Tests of the full LISA metrology baseline architecture can be performed with the BB. This is achieved by combining the capabilities for phase measurements, Time-Delay interferometry, clocktone transfer, testmass readout, laser control and inter spacecraft communications into a single device.

\begin{figure}
\begin{center}
\includegraphics[width=0.75\textwidth]{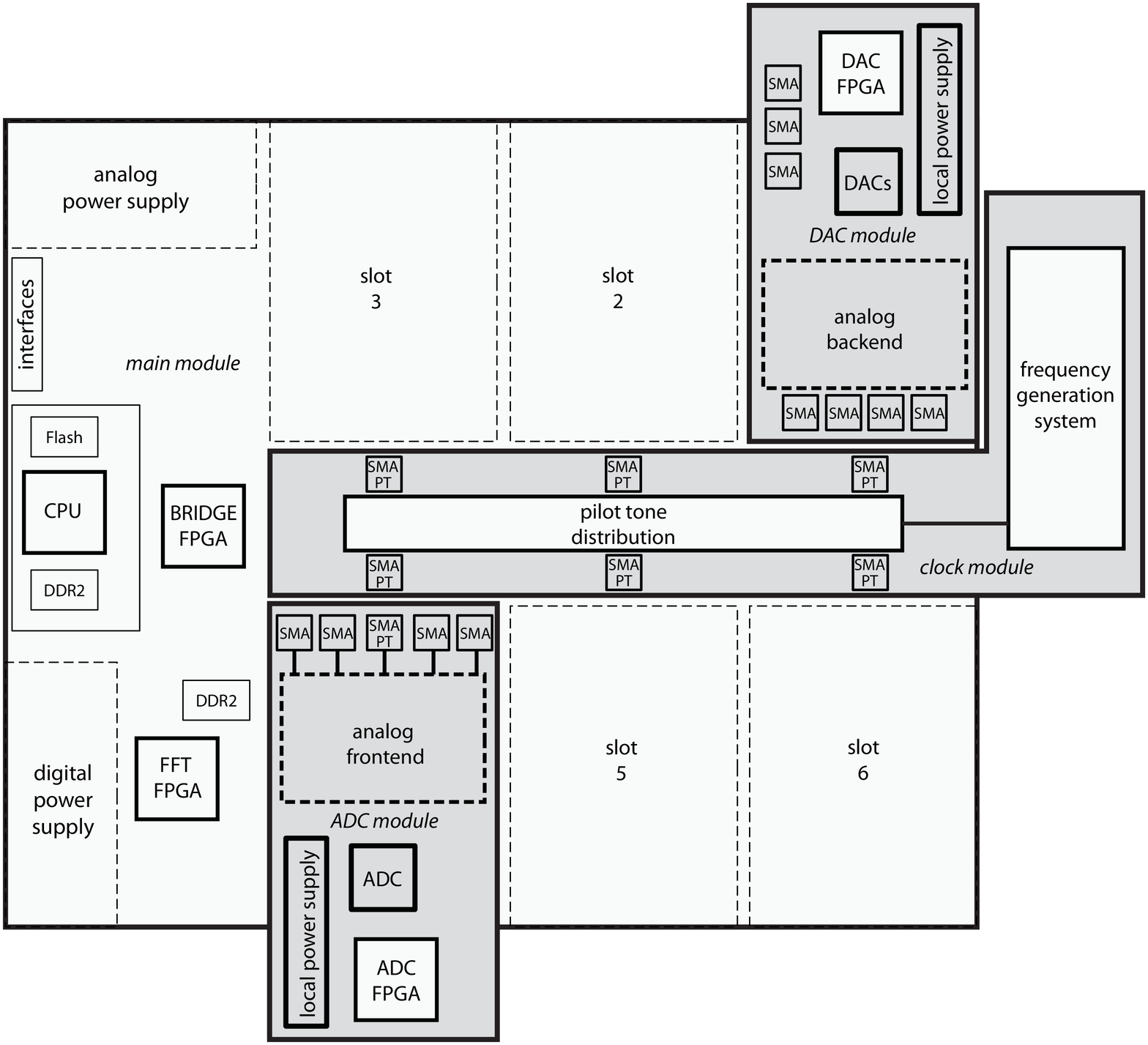}
\caption{\label{fig1} Schematic overview of the LISA phasemeter Breadboard model. The main module carries up to five ADC modules, one DAC module and one clock module. The pilot tone distribution is located in the middle of the BB to shorten its critical path. The FPGAs generating most of the heat in the system are located at the outside, allowing effective cooling and protection of the cricital analog circuits.}
\end{center}
\end{figure}

The BB (figure \ref{fig1} shows a schematic layout) consists of four types of modules, each containing different functionalities. The main module is the central part of the BB and connects to all other modules and the outside world. 
One ADC module handles the readout of up to four analog input channels and one BB can carry up to five modules, leading to a total number of 20 input channels.
The clock module contains the frequency generation system and the pilot tone distribution.
Furthermore each BB carries a DAC module, which produces all analog output signals of the phasemeter.
The modular approach was chosen to allow independent testing and optimization of critical components and to reduce production cost and risk. 
In the following we will give a detailed overview of each module.

\section{ADC module}

The actual measurement of the incoming signals is implemented on this module. The measurement chain consists of an analog frontend stage, preparing the signal for digitisation, a fast ADC, sampling with $80\,$MHz, and a FPGA containing the digital system processing. This includes the readout of the signal phase (\cite{Shaddock2006, Bykov2009}), the readout of inter-satellite ranging and communications (\cite{Esteban2011, Sutton2010}) and the readout of the clock tone sidebands (\cite{Heinzel2011}).

\subsection{Analog frontend}
The analog frontend includes a gain stage, an anti-aliasing filter, an addition of a pilot tone for ADC jitter correction and some signal conditioning for the ADC. Preinvestigations have shown that this is one of the most critical components of the BB, therefore we included the possibiliy of exchanging the frontend by soldering an alternative PCB board on top.

\subsection{Digital signal processing}
The actual phase readout of the MHz signals is performed by the use of specificly designed phase-locked loops (PLL). Each type of signal tone (science, pilot and sideband) is tracked seperately and the computed signals are decimated, downsampled and streamed to the further processing chain. The decimation rates and interface speeds are designed according to the bandwidth requirements of a potential DFACS system and the implementation of a digital laser lock. The readout of the inter-satellite ranging is performed in two delay-locked loops (DLL), which are tracking the local and remote pseudo-random noise (PRN) codes. The received delays and data are also transmitted to the CPU for storage and further processing.

\section{DAC Module}
The BB will also generate analog signals, which is done on the DAC module. It contains an FPGA, allowing to implement low-level computations and interfacing digital to analog converters and an analog signal conditioning, including additional gains and filtering. One of the main functionality is the generation and data encoding of the PRN signals, necessary for the intersatellite communication and data transfer. We have implemented three PRN outputs, to allow tests in a three laser setup. The other functionality is the generation of control signals for local lasers, to establish the LISA locking scheme. This includes the generation of four analog signals for controlling up to two slave lasers and the implementation of control algorithms in the FPGAs. This can either be the digital equivalent to an analog frequency offset phase-lock (\cite{Diekmann2009, Cruz2006}), or the implementation of more complex algorithms, like arm-locking (\cite{Yu2011}).

\section{Clock Module}
The clock module supplies the system clock and the pilot tone to the BB. It either generates or receives two GHz signals with a fixed phase relation to the pilot tone, which are used for the clock tone transfer. It consists of an analog signal chain and is divided into the generation of signals in the Frequency generation section and the distribution of the pilot tone to the ADC modules. The system sampling clock is distributed via the PCB board, while the pilot tone is seperated, to ensure its stability. The positioning of the clock module allows to optimize the signal distribution and allows to decrease the influence of strong heat dissipation by the FPGAs. Similar to the analog frontend of the ADC modules this component is a critical factor of the BB. It contains various testing possibilities to ensure the required funtionality and performance, as well as its own stabilised power supply.

\section{Main Module}

\begin{figure}
\begin{center}
			\begin{minipage}{0.4\linewidth}
			\includegraphics[width=1\textwidth]{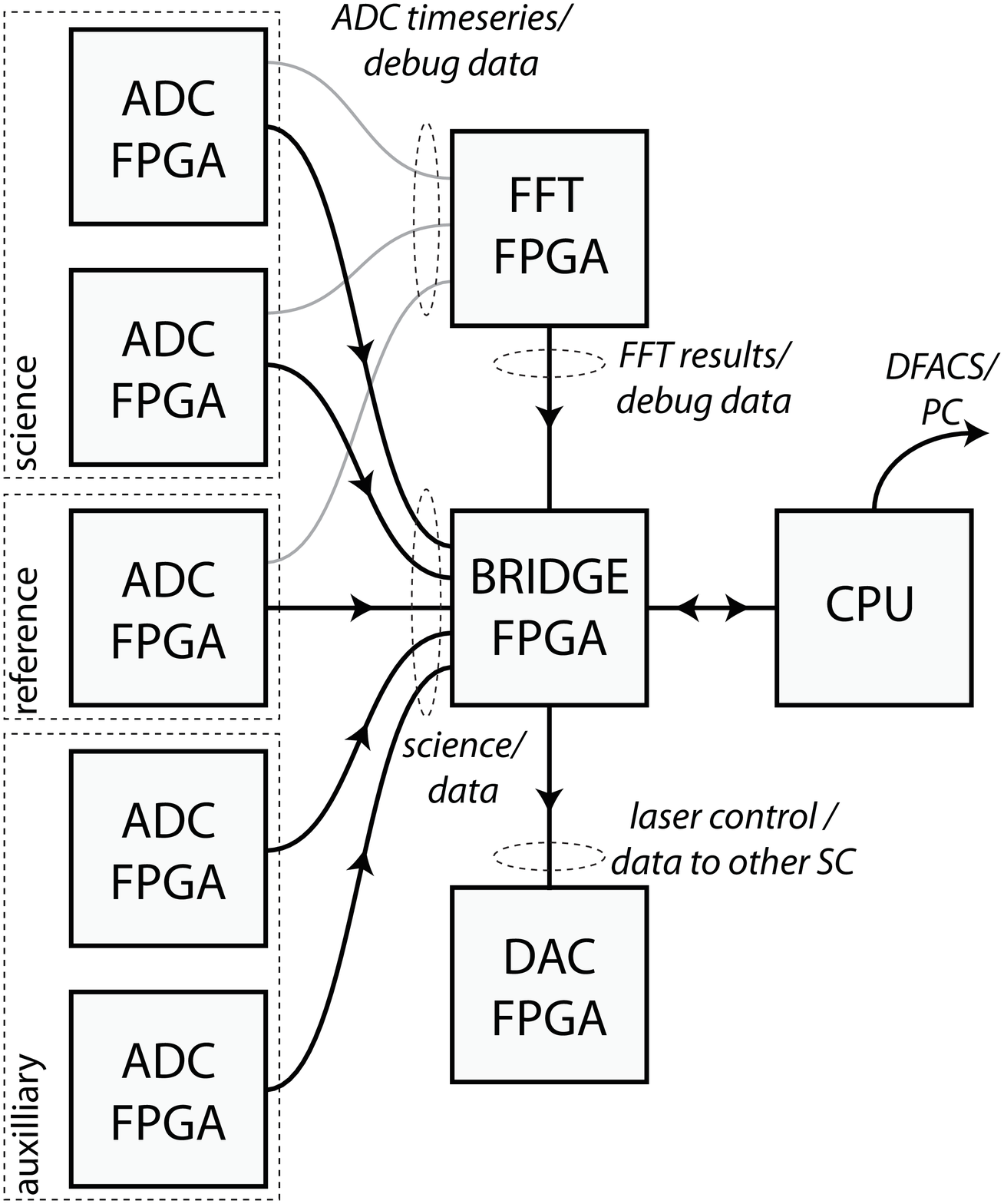}
			\end{minipage} 
			\hspace{0.05\linewidth}
			\begin{minipage}{0.3\linewidth}
			\includegraphics[width=1\linewidth]{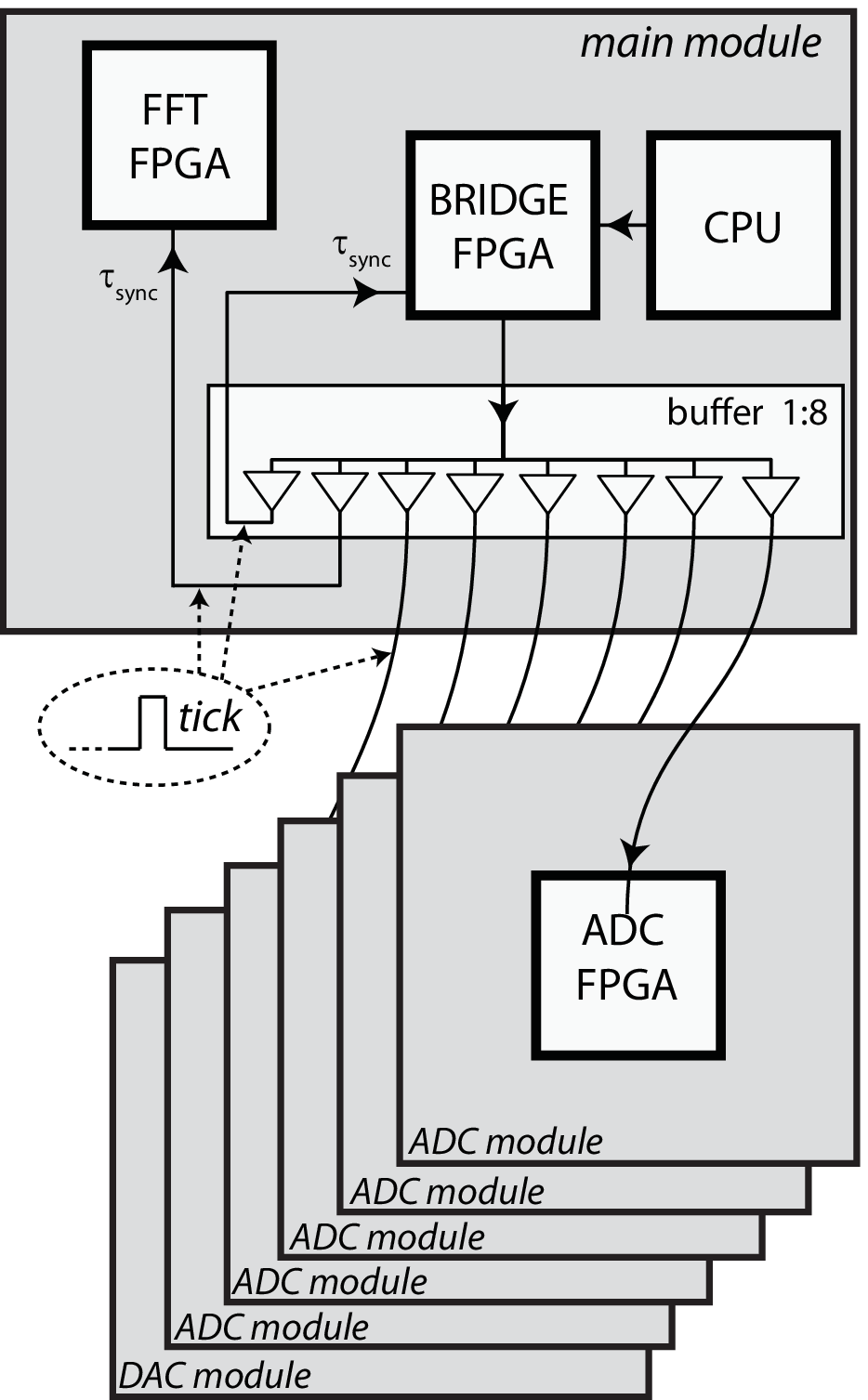}
			\end{minipage}
						\caption{\label{fig2} (left) System communication structure. All communication is routed through the bridge FPGA, with the exception of the highspeed data for the FFT FPGA. (right) Time tick distribution in the system via the bridge FPGA. All delays are matched, as well as the clock signals (not shown here), to allow a system wide hardware synchronization.}
	\end{center}
\end{figure}

Besides the function as baseplate and power supply, the main module contains central parts of the functionality. A separate "Bridge FPGA" is handling the system communication between the different FPGAs and a CPU, which is controlling the BB. A sketch of this is shown in figure \ref{fig2} (left). The CPU is also performing parts of the digital signal processing, data storage and the potential communication to a DFACS simulator. In addition the main board contains another FPGA, which is dedicated to perform Fast-Fourier Transforms of the incoming signal for lock acquisition procedures. \\

\subsection{Bridge FPGA}
The bridge FPGA is the central switch for most communications and is directly interfaced to the CPU, sending data to it and distributing commands from it to the respective submodules. For synchronising the various FPGAs system a time tick can be triggered at the bridge (see figure \ref{fig2}, (right)). This tick is sent to all FPGAs with the same time delay, also back to the bridge. This allows to synchronize clock counters in all systems, as well as any other action to a single clockcycle. Together with the pilot tone correction it forms the basis for meeting the stringent LISA timing requirements necessary for implementing TDI and clock noise corrections.
\\

\subsection{FFT FPGA}
To ensure a maximum acquisition performance the FFT FPGA can perform four full dutycycle FFTs with the complete ADC timeseries of an ADC module, transmitted by dedicated fast interfaces. It also hosts a peak detection algorithm and the possibility of storing a fullspeed datastream into dedicated DDR RAM. This stream is then read out via the CPU and can be used to perform debugging and signal analysis for short sequences at the full sampling rate.

\subsection{CPU}
The CPU will control the BB, perform housekeeping procedures, change operating modes and ensure synchronisation of all submodules. It is also performing several tasks for the digital signal processing. This includes a final data filtering and decimation to prepare the data for the DFACS system or for transmission to ground. The CPU is also handling the data storage and the preperation of data for the inter spacecraft communication.

\section{Outlook}
After the assembly of the BB, we will perform functionality and performance tests with electronic signals. The performance goals of all subsystems are calculated from the requirement of $1\,\textrm{pm}/\sqrt{\textrm{Hz}}$ in the range of $0.1\,\textrm{mHz}$ and $1\,\textrm{Hz}$ with the standard noise shape. Future steps are tests in LISA like optical setups and the evaluation of a design for space qualification.

\acknowledgements This work was funded by the European Space Agency (ESA) as part of the technology development for LISA. The authors greatfully acknowledge the support by the International Max-Planck Research School for Gravitational Wave Astronomy (IMPRS-GW), by QUEST (Centre for Quantum Engineering and Space-Time Research) and by Deutsches Zentrum f\"ur Luft-und Raumfahrt (DLR) (reference 50 OQ 0601).

\bibliography{gerberding_MyBibli}
\bibliographystyle{asp2010}

\end{document}